\documentclass{iopart}
\usepackage[latin1]{inputenc}
\usepackage[dvips]{graphicx}
\usepackage{graphicx}
\usepackage{bm}
\usepackage{epsfig}

\newcommand{\unit}[1]{\ensuremath{\,\mathrm{{#1}}}}
\newcommand{\+}{$^+$}

\begin{document}
\title{Formation of ultracold LiCs molecules}
\author{S D Kraft, P Staanum \footnote{Also at: Institut f\"ur Quantenoptik, Universit\"at Hannover,
Welfengarten 1, 30167 Hannover, Germany}, J Lange, L Vogel, R Wester
and M Weidem\"uller}

\address{Physikalisches Institut, Universit\"at Freiburg,
Hermann-Herder-Stra{\ss}e 3, 79104 Freiburg, Germany}
\ead{stephan.kraft@physik.uni-freiburg.de}

\begin{abstract}

We present the first observation of ultracold LiCs molecules. The
molecules are formed in a two-species magneto-optical trap and
detected by two-photon ionization and time-of-flight mass
spectrometry. The production rate coefficient is found to be in the
range $10^{-18}\unit{cm^3s^{-1}}$ to $10^{-16}\unit{cm^3s^{-1}}$, at
least an order of magnitude smaller than for other heteronuclear
diatomic molecules directly formed in a magneto-optical trap.

\end{abstract}

%%% ----------------------------------------------------------------------

\section*{Introduction}\label{sec:intro}

Owing to their permanent electric dipole moment, ultracold
heteronuclear dimers offer many exciting possibilities for studies
of interacting dipolar gases. At ultralow temperatures the
interaction energy of polar molecules in an electric field can
largely exceed their thermal energy. Hence processes such as elastic
and inelastic collisions can be manipulated by applying electric
fields \cite{Bohn2001, Ticknor2005, Avdeenkov2006, Krems2005}. On
the one hand, an array of cold polar molecules has been proposed to
represent a quantum computation device, where each qubit is defined
by the orientation of a molecular dipole relative to an external
electric field \cite{DeMille2002}. On the other hand, chemical
reactions of polar molecules at vanishing thermal energy have been
proposed to be controllable by suitable external electromagnetic
fields \cite{Krems2005, Avdeenkov2003}. Despite the existence of
potential energy barriers, such reactions feature significant rate
coefficients at low temperatures in the Wigner threshold regime
\cite{Balakrishnan2001, Bodo2002}.  Interesting complementary
trapping concepts for polar molecules are electrostatic
\cite{Meerakker-electrostatic-trapping, Rieger2005}, magnetic
\cite{Wang-KRb} and microwave traps \cite{DeMille2004}.
Experimentally, inelastic ultracold homonuclear atom-molecule
collisions have been studied in optical dipole traps
\cite{Staanum2006, Zahzam2006,Thalhammer2006}. Inelastic and elastic
collisions involving polar molecules are now open for similar
investigations offering new perspectives.

Large molecular electric dipole moments are required in order to
make experiments on dipole-dipole interactions and electric field
control of ultracold polar molecules feasible. At the current
state-of-the-art, heteronuclear alkali dimers seem to be
particularly suited. Such ultracold polar dimers can be formed by
photoassociation from ultracold samples of their atomic constituents
as demonstrated for RbCs \cite{Kerman-RbCs}, KRb \cite{Wang-KRb,
Mancini-KRbmolecules}, and NaCs \cite{Haimberger-groundstate}. The
dimer LiCs is predicted to have the largest dipole moment of about
5.5\, Debye for low lying vibrational states of the X$^1\Sigma^+$
electronic ground state \cite{Igel-Mann_1986,Aymar2005}. Recent
theoretical studies demonstrate the possibility for manipulating
Li-Cs collisions \cite{Krems2006}, rovibrational dynamics of LiCs
molecules \cite{Gonzalez-Ferez2006}, binary inelastic collisions
between KRb or RbCs \cite{Avdeenkov2006} and implementing the
mentioned quantum computation proposal using optically trapped KRb
or RbCs \cite{Kotochigova2006}.

Here we present the first observation of ultracold LiCs molecules.
These are formed out of an ultracold gas of Li and Cs atoms in two
spatially overlapped magneto-optical traps (MOTs). As observed in
several experiments, homonuclear
\cite{Gabbanini-Rbmolecules,Fioretti-Csmolecules,Takekoshi-OpticalTrappingCs2}
and heteronuclear \cite{Mancini-KRbmolecules,Haimberger-groundstate}
alkali dimers can be formed 'directly' in the MOT, i.e., without
applying a dedicated photoassociation laser. The molecule formation
is generally attributed to photoassociation induced by the trapping
laser light \cite{Caires2005}; three-body recombination has also
been conjectured as a possible cause for ultracold molecule
formation in a MOT \cite{Gabbanini-Rbmolecules}. In photoassociation
by the trapping light, a colliding atom pair is excited at long
range to a molecular potential which asymptotically correlates to
one excited atom and one ground state atom. Subsequently, the
excited molecule decays into two free atoms or into the molecular
electronic ground state, thus forming an ultracold dimer in the
latter process. According to the semi-classical model by Wang and
Stwalley \cite{Wang-Polar} the probability for the excitation step
is proportional to the reduced mass of the molecule to the power of
9/4, which yields a significant reduction for LiCs as compared,
e.g., to NaCs for which molecule formation rates around
$10-100\unit{s^{-1}}$ were observed in a MOT. In fact, light-induced
inelastic collisions between ground state Li and excited state Cs
atoms in a two-species MOT have been observed by our group as a
source of substantial atom loss already some years
ago~\cite{Schloeder}. The loss is due to the large kinetic energy
gained by moving inwards on the excited state potential curve
\cite{Gensemer1998} and possibly partly due to formation of LiCs
molecules. In this work, using a dedicated scheme for the detection
of LiCs molecules in the electronic ground state, we show that
ground state LiCs molecules are indeed formed in a two-species MOT
and we estimate their production rate.

\section*{Two-species magneto-optical trap}\label{sec:apparatus}

\begin{figure}[h]
\centering\includegraphics[width=9cm]{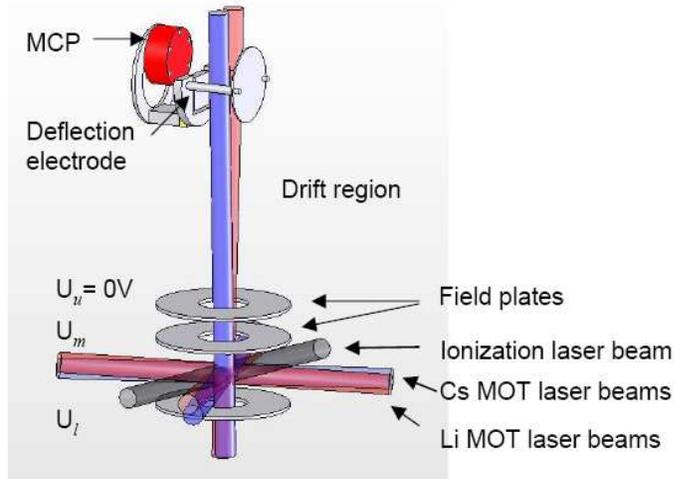}
\caption{\label{fig:setup} Sketch (not to scale) of the central part
of the two-species magneto-optical trap with time-of-flight mass
spectrometer of Wiley-McLaren type. Ions produced by photoionization
of trapped atoms and molecules between the two lower field plates
are accelerated upwards. After 30 cm they are deflected off the
vertical axis onto a microchannel plate detector (MCP).}
\end{figure}

The LiCs molecules are formed from Li and Cs atoms trapped in two
overlapped magneto-optical traps. The laser light for the Li MOT is
provided by two injection locked diode lasers (Philips, CQL-806),
one detuned 20\,MHz below the F=2 to F'=3 transition of the D2 line
for cooling, the other tuned to the F=1 to F'=2 transition for
repumping. The two beams are combined on a beamsplitter. Three
mutually perpendicular and retroreflected beams (4\,mW power in each
horizontal beam and twice the power in the vertical direction;
18\,mm diameter), derived from the combined beam, form the MOT. The
Cs MOT is formed by one retroreflected beam in the vertical
direction and four pairwise counterpropagating beams in the
horizontal plane. Each horizontal beam has a power of 4\,mW (twice
as much for the vertical beam) and a diameter of 16\,mm. The cooling
light is provided by a DBR laser diode (SDL-5722-H1), which is
locked 16\,MHz to the red of the F=4 to F'=5 transition near
852\,nm. A second diode laser (SDL-5712-H1) is used for repumping
from the F=3 state. For optimizing the spatial overlap of the MOTs,
the Li and Cs MOT laser beams are adjusted separately while
observing the fluorescence light of both MOTs in two perpendicular
planes using two CCD cameras. The central part of the experiment is
sketched in Fig.\,\ref{fig:setup}.

Both MOTs are loaded from Zeeman slowed atomic beams coming out of a
two-species oven, which consists of two separate chambers connected
by a capillary \cite{Stan-DoubleOven}. Each chamber can be heated
separately and hence the flux of the two atomic species can be
controlled independently. The front chamber containing Li is heated
to 350°C while the rear chamber containing Cs has a temperature of
150°C. The oven nozzle and the capillary are kept at 365°C to
prevent clogging. The Li (Cs) Zeeman slower laser is detuned 76\,MHz
(24\,MHz) below the cooling transition. Under these conditions we
achieve loading rates of $5\times10^7$\,s$^{-1}$ and
$1\times10^8$\,s$^{-1}$ for the Li and the Cs MOT, respectively.
Under steady-state conditions, $1.5\times10^8$ Li atoms and
$1.5\times10^8$ Cs atoms are trapped at a density of
$1\times10^{10}\unit{cm^{-3}}$ and $5\times10^9\unit{cm^{-3}}$,
respectively. Due to light-induced inelastic collisions
\cite{Schloeder}, the number of trapped atoms is reduced to 75\,\%
for Lithium as well as Cesium as compared to trapping only a single
species in the MOT.

\section*{Time-of-flight mass spectrometer}\label{sec:TOF}

To sensitively detect small numbers of molecules in an atomic gas,
pulsed two-photon ionization is combined with time-of-flight mass
spectrometry. The laser pulses are delivered by a pulsed dye laser
(Radiant Dyes Narrowscan with 7\,ns pulse width and 13\,mJ energy
per pulse). Based on ab initio potentials \cite{Korek_2000} and
previous spectroscopic studies in a heat pipe \cite{pashov} an
ionization wavelength range near 680\,nm was chosen to achieve
resonantly enhanced two-photon ionization of LiCs. The pulsed dye
laser beam is expanded to a $1.5\unit{cm^2}$ cross sectional area in
order to have a large ionization volume. The laser pulses produce a
significant atomic ion yield on the detector by non-resonantly
ionizing both Li and Cs atoms from the magneto-optical traps, either
with two photons from the 2p$_{3/2}$ and 6p$_{3/2}$ states,
respectively, or with three photons from the atomic ground states.
To separate a small signal of LiCs\+ from a large signal of Cs\+, a
high mass resolution is required.  For this purpose we implemented a
Wiley-McLaren time-of-flight mass spectrometer (see
Fig.\,\ref{fig:setup} and Refs.\,\cite{Kraft-TOF,Wiley-McLaren}).

Upon formation, all ions are accelerated upwards by applying
voltages to the field plates as indicated in Fig.\,\ref{fig:setup}.
After passing a 30\,cm drift region they are deflected onto a
microchannel plate detector (MCP) placed off the vertical axis in
order to allow optical access along this axis. By suitably adjusting
the ratio between the voltages on the lower (U$_l$) and the middle
(U$_m$) field plate as well as the voltage on the deflection
electrode (U$_{defl}$), the ions are time-focused onto the detector,
i.e., ions of the same mass arrive at the MCP within a narrow time
interval even though they are formed at different positions within
the ionization volume. The resulting voltage pulse from the MCP, due
to Cs\+ ions formed by ionization of atoms from the MOT, appears at
a time $\tau_{Cs}=12.74$\,$\mu$s after the laser pulse and is about
50\,ns wide. The Cs$_2$\+ ion peak is slightly broader (70\,ns),
since the molecular ions are produced in a much larger volume than
the ions originating from trapped Cs atoms. The mass difference
between LiCs\+ and Cs\+ of 5\% leads to a difference in time of
flight (scaling as $\sqrt{\rm{mass}}$) of only 2.5\%. We therefore
expect the LiCs\+ peak to appear at $\tau_{LiCs}=13.07$\,$\mu$s,
roughly 300\,ns after the Cs\+ peak, with a similar width as the
Cs$_2^+$ peak.

Since the Cs MOT contains many more atoms than LiCs molecules are
expected to be formed, it is necessary to suppress the Cs\+ signal
as much as possible. In order to avoid two-photon ionization of
Cs(6p$_{3/2}$) atoms we block the laser beams of the Cs MOT and of
the Cs Zeeman slower in a 2\,ms period before the dye laser pulse
arrives, using mechanical shutters. During this period all excited
Cs atoms decay to the electronic ground state and hence only
three-photon ionization of Cs(6s$_{1/2}$) is possible. With this
scheme we suppress the Cs\+ signal by a factor of 50, resulting in
about one detected Cs\+ ion per dye laser pulse.

\section*{Detection of LiCs molecules}\label{sec:results}

\begin{figure}%[p]
\centering\includegraphics[width=13cm]{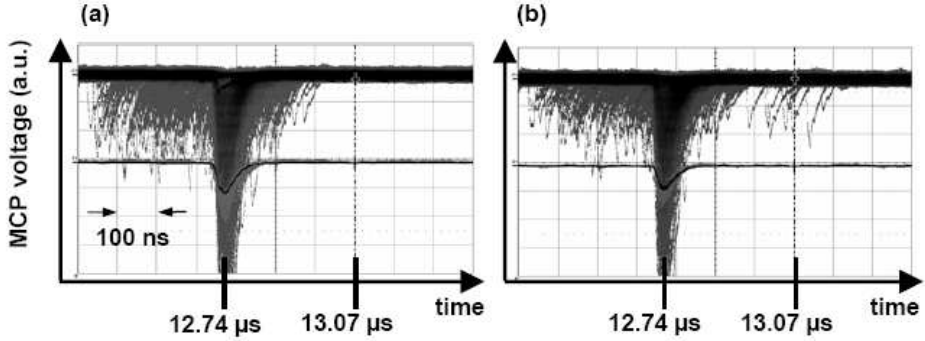}
\caption{\label{fig:TOFtrace} Oscilloscope images of time-of-flight
traces for photoionisation (a) in a single-species Cs MOT, and (b)
in a two-species Li-Cs MOT. The upper part of each image shows about
1100 time-of-flight traces on top of each other. The lower curve
represents an average of the traces. The Cs\+ ions arrive at
12.74\,$\mu$s, the LiCs\+ ions around 13.07\,$\mu$s.}
\end{figure}

In a first set of measurements we record time-of-flight traces with
the dye laser wavelength set to 682.78\,nm.
Fig.\,\ref{fig:TOFtrace}(a) shows an oscilloscope image recorded
with only the Cs MOT present, the laser beams of Li MOT and Li
Zeeman slower being blocked. In the upper part of the image the MCP
voltage of about 1100 time-of-flight traces laid on top of each
other is shown. The lower curve is an average over all traces. Every
time an ion hits the MCP detector a dip in the voltage is observed.
On both images we observe a large 50\,ns wide peak centered at
12.74\,$\mu$s originating from about one Cs\+ ion per dye laser
pulse, produced by non-resonant three-photon ionization of Cs MOT
atoms.

Additionally, we find a broad background of ions around the Cs\+
peak in a time interval from 300\,ns before to 200\,ns after the
Cs$^+$ arrival time. The detection rate of such background ions is
about one per ten dye pulses and therefore constitutes an important
source of background events near the expected LiCs\+ peak. These
ions can be attributed to fast Cs$^+$ ions formed by ionization of
fast Cs atoms: Weakly bound electronic ground state Cs$_2$ molecules
(formed by photoassociation induced by Cs MOT light) are
photodissociated by a first dye laser photon. After dissociation the
two atoms are either in the Cs(6s$_{1/2}$) and Cs(6p$_{3/2}$) or in
the Cs(6s$_{1/2}$) and Cs(6p$_{1/2}$) states. Depending on the state
they each have a kinetic energy of 0.18\,eV or 0.22\,eV,
respectively. The resulting Cs(6p) atom is then ionized by
two-photon ionization. Cs\+ ions emerging from this process have a
velocity of about 500\,m/s and 550\,m/s, respectively, which leads
to a broad distribution as observed in the measured time-of-flight
trace. To avoid the fast Cs\+ ion background at the expected arrival
time of the LiCs\+ ions, we reduced its width to $\Delta
\tau=500\unit{ns}$ by applying large absolute voltages U$_l$ and
U$_m$ to the field plates, which in turn decreases the ratio
$\frac{\Delta\tau}{\tau_{LiCs}\,-\,\tau_{Cs}}$.

In addition, the arrival time of the fast ions can be shifted to
shorter arrival times relative to the narrow Cs\+ peak by
adjusting the ratio of the field plate voltages (U$_l$/U$_m$) at
the cost of a slightly non-optimal time-focussing. By applying
these fine-adjustments to the field plates, we can clearly
separate the fast Cs\+ ion background from ions at the expected
LiCs\+ arrival time. More specifically, at U$_l$=800\,V,
U$_m$=610\,V and U$_{defl}$=1990\,V the rate of Cs\+ ions
appearing within 100\,ns of the expected LiCs\+ arrival time at
$13.07\unit{\mu s}$ is about one ion per 4500 dye pulses.

In Fig.\,\ref{fig:TOFtrace}(b), recorded with both MOTs being
present, we observe several ions in excess of the Cs\+ background,
which impact on the detector at the expected LiCs\+ arrival time, in
addition to the Cs\+ peak and a broad background as in
Fig.\,\ref{fig:TOFtrace}(a). These ions are only observed when both
the Li and the Cs MOTs are present. In a window of 100\,ns width
around the expected LiCs\+ arrival time four ion signals are counted
in 1100 dye pulses. To exclude that these ions are produced by
processes induced by the Li trapping light at 670\,nm, such as
photodissociation of Cs$_2$, we compared the measurements in
Fig.\,\ref{fig:TOFtrace}(a), where all Li lasers are blocked, to
measurements with the Li MOT laser beams present but the Li Zeeman
slower laser beam being blocked, i.e. with no Li atoms being loaded
into the MOT. In both cases only about one ion every 4500 cycles
appears within a 100\,ns window at the expected LiCs$^+$ arrival
time. Thus, we conclude that the ions observed in
Fig.\,\ref{fig:TOFtrace}(b) at a much larger rate originate from
cold LiCs molecules formed out of cold Li and Cs atoms in the
two-species MOT.

Photoassociation of LiCs molecules induced by the laser light can
only occur via the Li(2s$_{1/2}$)+Cs(6p$_{3/2}$) asymptote.
Photoassociation to potentials correlating to the
Li(2p$_{3/2}$)+Cs(6s$_{1/2}$) asymptote is excluded since all
potentials are repulsive beyond 50\,\AA, with the exception of one
potential curve which is slightly attractive and supports a
potential minimum around 70\,\AA. However, this potential well is
extremely shallow so that photoassociation into this well is not
possible since the Li laser detunings are much larger than the well
depth \cite{Bussery1987}. Since the Cs MOT and Zeeman slower laser
light is blocked in a 2\,ms period before the ionization pulse, all
molecules formed via the Li(2s$_{1/2}$)+Cs(6p$_{3/2}$) asymptote
must have decayed to the electronic ground state before ionization.
In the case of LiCs formation by three-body recombination, which is
unlikely due to the comparatively small pair densities, the
molecules would be formed directly in the electronic ground state.
In either case, the detected LiCs\+ ions arise from ultracold LiCs
molecules in the electronic ground state.

The 2\,ms time interval between the blocking of the Cs MOT lasers
and the ionization pulse gives an upper bound on the velocity of the
molecules. Only molecules can be ionized which have not left the
ionization region resulting in a maximal velocity of 5 m/s. However,
the corresponding temperature of 0.3\,K can only be seen as an upper
bound. Based on other experiments with KRb and NaCs
\cite{Mancini-KRbmolecules, Haimberger-groundstate} we expect a
temperature for  LiCs on the order of 100\,$\mu$K given by the Cs
temperature.

Within the measured 1100 dye laser pulses, we detect
$(3.6\pm2.0)\times10^{-3}$ LiCs ground state molecules per
ionization pulse after background subtraction and assuming
Poissonian statistical uncertainties. In order to estimate the LiCs
production rate several factors must be taken into account. Due to
gravity  and finite kinetic energy the molecules leave the
ionization volume. Therefore, they can be ionized only in an
interval of 13\,ms after their production. Due to the 10\,Hz
repetition rate of the ion detection, only $13\%$ of the produced
molecules are available for ionization. Additionally, our detector
is estimated to have a detection efficiency of about $20\%$
\cite{Kraft-TOF}. Assuming that all molecules within the laser field
of the dye pulse are ionized, we derive a production rate of
$0.14\pm0.08$ molecular ions per second. As the ionization process
has not yet been investigated, the ionization probability is
estimated to lie between $10\%$ and $0.1\%$, depending on the
proximity of the dye laser wavelength to an intermediate resonance
state. This leads us to an estimate of the actual molecule
production rate between $1.4\pm0.8$ and $140\pm80$ LiCs molecules
per second.

In our experiments the Li MOT is larger than the Cs MOT. In this
case the molecular production rate can be approximately written as
\begin{equation*}
dN_{LiCs}/dt =\beta_{LiCs} N_{Cs} n_{Li}
\end{equation*}
where $N_{Cs}$ is the total number of Cs atoms, $n_{Li}$ the Li peak
density and $\beta_{LiCs}$ the rate coefficient of LiCs molecule
production. With typical values of our double MOT setup of $n_{Li}=
10^{10}\unit{cm^{-3}}$ and $N_{Cs}= 1.5\times10^{8}$ this leads to a
molecule production rate coefficient between
$10^{-18}\unit{cm^3s^{-1}}$ and $10^{-16}\unit{cm^3s^{-1}}$ for
$10\%$ and $0.1\%$ ionization probability, respectively. A similar
range of production rate coefficients results from a second set of
measurements, which we performed with a smaller Li MOT at a dye
laser wavelength of 681.08\,nm. The obtained range of rate
coefficients is in agreement with the calculated Franck-Condon
factors for polar alkali dimer photoassociation \cite{Wang-Polar}.
This calculation predicts for LiCs a rate coefficient of about an
order of magnitude smaller as compared to the formation rate of NaCs
measured as $10^{-15}\unit{cm^3s^{-1}}$ in a MOT
\cite{Haimberger-groundstate}. Both the LiCs and the NaCs formation
rates are much smaller than the observed rate for photoassociation
of KRb in the trapping light, measured to be
$8\times10^{-12}\unit{cm^3s^{-1}}$ \cite{Mancini-KRbmolecules}, due
to the much larger Franck-Condon factor in the KRb system.

\section*{Conclusion}\label{sec:conclusion}

We report on the first detection of ultracold LiCs molecules in the
electronic ground state. The molecule formation rate coefficient is
estimated to be between $10^{-18}\unit{cm^3s^{-1}}$ and
$10^{-16}\unit{cm^3s^{-1}}$. This coefficient is smaller than the
values obtained for NaCs and KRb \cite{Mancini-KRbmolecules,
Haimberger-groundstate}, which can be essentially explained by the
small reduced mass of LiCs and the very large $C_6$ coefficients for
KRb \cite{Wang-Polar,Bussery1987}. Future experiments on LiCs will
therefore include photoassociation with a tunable cw Ti:Sapphire
laser either in the two-species MOT or an optical dipole trap
\cite{mudrich2002:prl}, which should lead to LiCs production rates
larger by orders of magnitude \cite{Azizi2004}. Furthermore,
ionization spectra as a function of the dye laser wavelength will be
measured to analyze the vibrational state distribution of the formed
molecules. By transferring the formed molecules to low lying
vibrational states of the X$^1\Sigma^+$ state by stimulated Raman
processes \cite{Sage-RbCs-v0} or using microwave fields, cold polar
molecules with electric dipole moment as large as 5.5\,Debye
\cite{Aymar2005} can be obtained for studies on dipolar gases and
ultracold chemistry.

This work was supported by the Deutsche Forschungsgemeinschaft in the frame of
the Schwerpunktprogramm 1116 and by the European Commission in the frame of
the Cold Molecule Research Training Network under contract HPRN-CT-2002-00290.

\end{document}